# Scaling behavior of entropy estimates

Thomas Schürmann

Germaniastraße 8, 40223 Düsseldorf, Germany

Email: t.schurmann@icloud.com

**Abstract.** Entropy estimation of information sources is highly non trivial for symbol sequences with strong long-range correlations. The rabbit sequence, related to the symbolic dynamics of the nonlinear circle map at the critical point as well as the logistic map at the Feigenbaum point have been argued to exhibit long memory tails. For both dynamical systems the scaling behavior of the block entropy of order *n* has been shown to increase like $\propto \log n$. In contrast to probabilistic concepts, we investigate the scaling behavior of certain non-probabilistic entropy estimation schemes suggested by Lempel and Ziv in the context of algorithmic complexity and data compression. These are applied in a sequential manner with the scaling variable being the length *N* of the sequence. We determine the scaling law for the Lempel-Ziv entropy estimate applied to the case of the critical circle map and the logistic map at the Feigenbaum point in a binary partition.

1. Introduction

Partially random chains of symbols $s_1, s_2, ..., s_n$, drawn from some finite alphabet, appear in practically all sciences. Examples are human written texts, DNA sequences, spins in 1-dimensional magnets, cellular automaton as well as bits in the storage and transmission of digital data like music scores and pictures. Within this context it would be interesting to determine to what degree these sequences can be "compressed" without losing any information. This question was first posed by Shannon from a probabilistic point of view [1], who showed that the respective measure in this context is the entropy (or average information content) $h$. In the context of statistical physics, $h$ is related to the thermodynamic (Boltzmann) entropy of the underlying system. One reason for the significance of Shannon's framework is its association with the achievable compression ratio in the limit of very long sequences.

Formally an information source is by definition a mechanism that produces messages over a finite alphabet $A = \{0,1,...,d-1\}$, i.e. a set of symbols which will be of size $d$. A message of length $n$, i.e. the symbol sequence under consideration shall be denoted $s_1^n = s_1, s_2,..., s_n$, with $s_i \in A$. A code $C$ is a translation mechanism (an algorithm) that, for each $n$, takes as input a message from $A^n$ and transforms it into a binary sequence. Such a transformation is thus a fixed-length to variable-length encoding. Assuming the message to be produced by a stationary and ergodic source, then the quantity of importance in determining the entropy rate $h$ of the source is the block entropy

$$H_n = -\sum_{s_1,...,s_n} p(s_1^n) \log p(s_1^n), \qquad (1)$$

where $p(s_1^n)$ are the probabilities of subsequences or "words" of fixed length $n$, and the logarithm is of base two in the following. The Shannon entropy (or the entropy of the source) for a stationary and ergodic process is formally given by

$$h = \lim_{n \to \infty} \frac{H_n}{n}. \qquad (2)$$

Now, the most fundamental theorem of information theory is due to Shannon and asserts that *any code has an expected length per symbol that is at least as large as the entropy rate (2) of the source.* For the numerical estimation of the entropy, one can do simple word counting hence, achieving a relative distribution for subsequences or words of a fixed length $n$. In the limit of large data sets, the relative frequency distribution yields the underlying probability distribution.

The estimation of entropy can be highly non-trivial depending on the complexity of the sources and this is always true in the case if there are strong long-range correlations. Although such correlations can help to achieve higher compression rates (since they reduce the uncertainty of yet unseen symbols). But finding them and taking them into account can be very difficult because of the exponential increase of the number of different blocks or "words" with the length $n$ of the blocks. Several works on the estimation of $h$ and finite sample corrections for the block entropy $H_n$ have been discussed in literature[4,5,6]. Empirically investigations lead to power-law scaling, $H_n \approx a n^\mu + b$, $\mu \approx 0.5$, in the case of certain samples of texts and $\mu \approx 0.25$ for some classical music[5,7]. For Markov processes of order $m$, one easily gets that the entropy of the source $h$ is reached for blocks of length $n = m$.

Also of special interest is the border between periodicity and chaos, when the dynamics of the logistic map approaches chaotic in the way of period doubling. At the Feigenbaum point one finds for blocks of length $n = 2^k$, a result obtained by Grassberger [8,9],

$$H_n \propto \log(n), \qquad (3)$$

i.e. the entropy per symbol decays very slowly $\propto \log(n)/n$ to zero. Another example of long-range order can be viewed as the symbolic dynamics associated with the sine circle map at the critical point [10,11]. As in the case of the logistic map the scaling behavior of the block entropy is $\propto \log(n)$ [12].

However, a second famous theorem of Shannon asserts that *the lower bound (2) is asymptotically achievable*. This leaves plenty room for algorithmic design. As matter of fact, coding algorithms separate into two groups: first codes that are designed for a specific (known) probability distribution over the input string. Or second, universal codes that do not assume a probabilistic distribution to be known *a priori* and do their best to come close to the optimum over an entire class of sources. Amongst the first group, one find Huffmann- and Shannon-Fano codes. Amongst the second group, the best known algorithms are the ones due to Lempel and Ziv [13,14,18]. Originally constructed to provide a complexity measure for individual finite sequences, the Lempel-Ziv (LZ) algorithm is similar in the spirit of the algorithmic complexity of Kolmogorov, Solomonoff and Chaitin [15-17]. However, it was shown [18] that, in the case of statistically stationary strings, it converges to the Shannon entropy when the length of the string tends to infinity.

## 2. Lempel-Ziv coding

Attempt to eliminate probabilistic ideas, the method with best chances of taking long-range correlations into account has been the coding schemes of Lempel and Ziv [13,14]. In this scheme the sequence of length $N$ is parsed into words of variable length such that the next word is the shortest word not seen in the past. The corresponding code consists of pairs of numbers: each pair being a pointer to the previous occurrence of the prefix of the phrase and the last bit of the phrase.

Formally, the sequence is broken into words $w_1, w_2, ...$ such that $w_1 = s_1$, and $w_{k+1}$ is the shortest new word immediately following $w_k$. For instance, the sequence $S = 10110101101110...$ is broken into segments

$$(1)(0)(11)(01)(011)(0110)(....$$

In this way each word $w_k$ with $k > 1$ is an extension of some $w_j$ with $j < k$ by one single symbol $s' \in A$. Any element of such partition is then simply encoded by the pair $(j, s')$. Hereafter, we consider only a binary alphabet, but an extension to any finite alphabet is straightforward. It is obvious that this is a good encoding in the sense that the string can be uniquely decoded from the code sequence. Both the encoder and the decoder built up the same dictionary of words, and thus the decoder can always find and add the new word. The encoding is efficient because for sequences of low entropy there are strong repetitions, such that the average length of the words $w_k$ increase faster, and the number of needed pairs $(j, s')$ slower, than for sources of higher entropy.

This Ziv-Lempel (ZL) coding is indeed a simplification of an earlier scheme by Lempel and Ziv [13], called LZ coding in the following. There the string is also broken up into a chain of words $w_1, w_2, ...$, but a word $w_k$ is there not necessarily an extension of a previous word $w_j$. Instead it can be an extension of any substring of $S$ which starts before $w_k$ (and maybe overlaps it). In the above example we get the different parsing

$$(1)(0)(11)(010)(11011)(0...$$

This seems more efficient than ZL parsing in the sense that the average word length increases faster and the algorithm can make better use of long-range correlation. Also, the code length per word is slightly larger, so that it is not clear whether its compression performance is indeed superior to ZL for small $N$. In any case the convergence of the compression rate with $N$ is not well understood theoretically in either of these schemes. More precisely, for both LZ and the ZL schemes, the entropy of stationary and ergodic sequences is related to the length $l_N$ of the associated encoding of previous words in the case of ZL, and of arbitrary previously seen strings for LZ via

$$h = \lim_{N \to \infty} \frac{l_N}{N}. \tag{4}$$

In this way, both parsing schemes are universal in the sense that they reach asymptotically the entropy of the source. As in the case of (2), i.e. the block entropy rate versus the block length *n*, the convergence of the code length (4) versus the length *N* of the sequence is from above. In the case of block entropies, the correlations are taken more and more into account when the length of the blocks become large, so that the information per symbol decreases with *n*. In the case of Lempel and Ziv, on the other hand, in addition to the specific sequence also the information about the probability distribution has implicitly to be encoded. This contributes mostly at the beginning, whence the information per symbol is highest at the beginning. Otherwise stated, the LZ encoding is self-learning, and its efficiency increases with the length of the sequence.

Slightly modifications of Lempel and Ziv's coding (i.e. prefix tree's, dictionary extension etc.) have been applied to entropy estimation of English texts by Grassberger [19] and others [20,21]. To get a deeper insight to the finite sample behavior, in the following we compute the rate of convergence associated with the estimate

$$h_N = \frac{l_N}{N}. \tag{5}$$

When looking at the case of memoryless (Bernoulli) sources with symbol "0" occurring with probability $p > 0$ and symbol "1" appearing with probability $q = 1 - p$, then the average excess of $h_N$ with respect to the entropy of the source associated with the ZL scheme is of order $\propto 1/\log N$ [22], whereas the corresponding expression for the LZ coding is known to be of order $\propto \log\log N / \log N$ and this is conjectured to be the right order. Therefore, assuming this conjecture, the ZL algorithm – based on "shorter" words in the parsing– is more efficient in the case of Bernoulli sources. Nevertheless, Shields [23] proved that such redundancy rates cannot be achieved for general stationary and ergodic sources. Furthermore, it must be observed that in most practical cases the above redundancy rates are not achievable.

### 3. Parsing the rabbit sequence

In the following the critical circle map is represented by partitioning the time series on a binary (generating) partition. Denoting the pieces of the partition by symbols "0" and "1", the dynamical system is mapped unique on a infinite binary string. This string, also called rabbit sequence [11] is generated by the grammar rule [12]

$$b_0 = 0,$$
$$b_1 = 1,$$
$$b_{r+1} = b_r b_{r-1}, \qquad \text{for } r \geq 1. \qquad (6)$$

The right hand side of the last equation formally represents the concatenation of both finite successors $b_{r-2}$ and $b_{r-1}$ of $b_{r+1}$, called fibonacci words. Thus, the rabbit sequence is equal to the infinite fibonacci word. The first few fibonacci words are

$$b_1 = 1$$
$$b_2 = 10$$
$$b_3 = 101$$
$$b_4 = 10110$$
$$b_5 = 10110101$$
$$b_6 = 1.0.11.010.11011.0$$

and the dots in the last word indicate the first words of the LZ incremental parsing scheme. Many interesting properties of fibonacci words concerning their symmetry, divisibility or properties of self similarity can be found in [11,24-26] and references therein. One property is that the length (i.e. the number of bits) of the $r$th fibonacci word, $b_r$, equals the $(r+1)$th fibonacci number $F_{r+1}$. The fibonacci numbers are defined by the recursive relation

$$F_0 = 0, \quad F_1 = 1,$$
$$F_{r+1} = F_r + F_{r-1}, \tag{7}$$

i.e. the first few are $0,1,1,2,3,5,8,13,...$ .

To determine $h_N$, we construct the incremental parsing associated with LZ algorithm. In the following let $s_i^j$ be the substring of $S$ starting at position $i$ and ending at $j$, $s_i^j = s_i, s_{i+1},..., s_j$. Then we have

**Theorem 1.** (Parsing)

Let $S$ be the rabbit sequence defined by (6). Then, for all $k \geq 1$, the incremental parsing of $S$, associated with the LZ-coding scheme is represented by the words
$$w_k = S_{F_{k+1}}^{F_{k+2}-1}. \tag{8}$$

**Proof.** First note that for all $k \geq 1$, the length of the $k$th word $w_k$ is equal to $F_k$, and the length of the concatenation of the first $M$ words, $w_1,..., w_M$, is $\sum_{k=1}^{M} F_k = F_{M+2} - 1$. Hence the number of words in any fibonacci word $b_r$ is $r-1$, and it follows that the length of the concatenation $w_1,..., w_{r-1}$ is just one symbol smaller than $b_r$. Let $u_r$ be that last symbol (or word of length 1) of $b_r$, then we get the following expression for the $r$th fibonacci word
$$b_r = w_1 w_2 ... w_{r-1} u_r. \tag{9}$$

**Lemma 1.** For $r \geq 2$, the last symbol of $b_r$ is
$$u_r = \frac{1-(-1)^r}{2}. \tag{10}$$

**Proof.** (by induction)

(i) Let $r = 2k$ be even, $k \geq 1$. For $k = 1$ it follows that $b_2 = b_1 0$. Step from $k$ to $k+1$: by induction assumption the last symbol of $b_{2k}$ is "0" and by definition (6) we have $b_{2(k+1)} = b_{2k+1} b_{2k}$. Hence, the last symbol of $b_{2(k+1)}$ is "0".

(ii) Let $r = 2k+1$ be odd, $k \geq 1$. For $k = 1$ it follows, $b_3 = b_2 1$, by definition. Step from $k$ to $k+1$: by assumption the last symbol of $b_{2k+1}$ is "1". By the rule (6) we have $b_{2(k+1)+1} = b_{2k+2} b_{2k+1}$. Hence, the last symbol of $b_{2(k+1)+1}$ is "1". (End of proof Lemma 1)

***Lemma 2.*** (dynamical phrase generation)

For $k \geq 1$, let $w_k^-$ denote the $k$th word of (8), but inverting the last symbol. Then the following rule of recursion holds:

$$w_1 = 1, \quad w_2 = 0,$$
$$w_k = w_{k-2} w_{k-1}^-, \quad \text{for } k \geq 3. \tag{11}$$

***Proof.*** Inserting expression (9) into the recursion relation of the fibonacci words (6) and using the identity $u_{k-1} = u_{k+1}$, it follows that

$$w_k = u_k w_1 w_2 ... w_{k-2}, \quad \text{for } k \geq 1. \tag{12}$$

Then, (11) will be proofed by induction:

*(i)* Let $k = 3$, then $w_3 = 11 = w_1 0^-$. Induction step: by assumption, (11) is true for $k$ fixed. Multiplying (11) by $w_{k-1}$ from left and using (12) for $k+1$, one gets $w_{k-1} w_k = w_{k+1}^-$. Multiplying the latter equation by $u_k w_1 w_2 ... w_{k-2}$ from left and using $u_{k+2} = u_k$, we get $w_{k+2} = w_k w_{k+1}^-$, which is the first part of the proof.

*(ii)* Let $k = 4$, then $w_4 = 010 = w_2 w_3^-$. Induction step: by similar arguments as in *(i)*, from $w_{k+1} = w_{(k+1)-2} w_{(k+1)-1}^-$, we get the final expression $w_{k+3} = w_{k+1} w_{k+2}^-$ which yield the desired result.

(End of proof Lemma 2)

Now, since the last symbol of successor $w_{k-1}^-$ in (11) is "flipped", it is sure that no extension to the right of any word $w_k$ can occur. To close up the proof of Theorem 1, we finally have to check that there is no substring equal to $w_k$ before $w_{k-2}$, i.e. before position $F_{k-1}$ in the rabbit sequence. But this property is given by Theorem (a) in [12] by setting $k = s+2$, i.e. all substrings of length $F_k - 1$, that start at position $1,..., F_k$ are mutually different.

The compression ratio (5) can be determined by computing the number of bits $l_N$, needed to encode the first $N$ symbols of the rabbit sequence. From Theorem 1 we know that the length of the $k$th word in the LZ parsing is equal to $F_k$. From Lemma 2 it follows that the reference word in the history of word $w_k$ is at the position equal to $F_{k-1}$. From the analytical expression of fibonacci numbers we get the approximation

$$F_k \approx \frac{\phi^k}{\sqrt{5}}, \quad \phi = \frac{1+\sqrt{5}}{2} = 1.618...,$$

which becomes exact when rounded to the nearest integer [28]. Since one needs approximately $\lceil \log i \rceil$ digits to encode an integer $i$, the code length of the $k$th word is equal $k \log \phi + O(1)$. Summing up these contributions we find

$$l_N = \frac{\log \phi}{2} M^2 + O(M). \tag{13}$$

The number of words, $M$, in the LZ-parsing is related to the number of symbols of the string by

$$N = F_{M+2} - 1. \tag{14}$$

Since, the inverse relation is (up to rounding errors)

$$M = \frac{1/2 \, \log 5 + \log(N+1)}{\log \phi} - 2. \tag{15}$$

Inserting the last expression into (13), the leading term of the estimate $h_N$ is

$$h_N = \frac{(\log N)^2}{2N \log \phi} + O(\frac{\log N}{N}). \tag{16}$$

## 4. Incremental parsing at the Feigenbaum point

Also of special interest is the border between periodicity and chaos, when the dynamics of the logistic map approaches chaotic in the way of period doubling. Various properties of the associated symbolic dynamics have been discussed in the context of block entropy computations [8,9,28]. The recursive grammatical rule at the Feigenbaum point is [28]

$$\begin{aligned} a_0 &= 1, \\ a_1 &= 10, \\ a_{k+1} &= a_k a_{k-1} a_{k-1}, \end{aligned} \tag{17}$$

for $k \geq 1$. The first few "Feigenbaum words" are

$$\begin{aligned} a_0 &= 1 \\ a_1 &= 10 \\ a_2 &= 1011 \\ a_3 &= 10111010 \\ a_4 &= 1.0.11.1010.10111011. \end{aligned}$$

The dots in $a_4$ indicate the first words of the LZ incremental parsing scheme. In contrast to the case of the fibonacci words, here the dots seem to be in coincidence with the end of any preceding $a_k$. More precisely this comfortable situation is stated in

*Theorem 2.* (Parsing)
Let $S$ be the symbol sequence generated by fibonacci words (17). Then the words in the LZ incremental parsing are

$$w_0 = 1,$$

$$w_k = a^-_{k-1}, \qquad \text{for } k \geq 1. \tag{18}$$

*Proof.* By definition of LZ, for $k=0$, we simply get $w_0 = 1$. For the case $k \geq 1$ we state the following

**Lemma 1.** For any $k \geq 1$ it is

$$a_{k-1}a_{k-1} = a^-_k. \tag{19}$$

*Proof.* The proof is by induction. Start of induction, $k=1$: it follows by definition that $a_0 a_0 = 11 = 10^- = a^-_1$. Induction step from $k$ to $k+1$: by induction assumption we have $a^-_k = a_{k-1}a_{k-1}$. "Flipping" the last bit in the equation it simply follows that $a_k = a_{k-1}a^-_{k-1}$. Multiplying by $a_k$ from left we get by definition (17), $a_k a_k = a_k a_{k-1} a^-_{k-1} = (a_{k+1})^- = a^-_{k+1}$. (End of proof Lemma 1).

By Lemma 1, for $k \geq 1$, any Feigenbaum word, i.e. the first $2^k$ bits of $S$ can be expressed as $a_k = a_{k-1}a^-_{k-1}$. Since the suffix $a^-_{k-1}$ is identical to the prefix $a_{k-1}$ except by the last symbol and bot are of equal length, it follows that $w_k = a^-_{k-1}$. Then, the preceding reference word of any $w_k$ starts at the beginning of $S$ and stops at the position of $w_k$ in $S$. (End of proof Theorem 2)

Thus, for $k \geq 1$, the length of $w_k$ is $2^{k-1}$, and the number of bits necessary for encoding $w_k$ is $k + O(1)$. Summing up these contributions we find the codelength

$$l_N = \frac{M^2}{2} + O(M). \tag{19}$$

The number of words, *M,* in the LZ-parsing is simply related to the number of symbols of the string (up to rounding errors) by

$$M = 1 + \log N. \tag{20}$$

Inserting the last expression into (19), we find

$$h_N = \frac{(\log N)^2}{2N} + O(\frac{\log N}{N}). \tag{21}$$

## 5. Conclusion

For the case of memoryless random sources the expected excess of $h_N$ over the entropy of the source associated with the ZL scheme is known to be of order $\propto 1/\log N$ [22], whereas for the LZ coding it is $\propto \log\log N / \log N$. Hence, in the case of Bernoulli sources, the ZL algorithm – based on "shorter" words in the parsing – is faster. On the other hand, in the case of the rabbit sequence or the logistic map at the Feigenaum point, the convergence of the LZ scheme is faster than for the memoryless case.

The author did not proof similar results for the ZL-parsing, but we conject the convergence of leading order $\propto \log N / \sqrt{N}$ for both dynamical systems, which is quite slower compared to LZ. Since, for sequences with long-range correlations one would expect a faster convergence of the LZ-scheme. Whereas for the case of Markov- or finite state sources it seems opposite.